# Effects of Dielectric Stoichiometry on the Photoluminescence Properties of Encapsulated WSe₂ Monolayers

*Javier Martín-Sánchez[1*], Antonio Mariscal[2], Marta De Luca[3], Aitana Tarazaga Martín-Luengo[1], Georg Gramse[4], Alma Halilovic[1], Rosalía Serna[2], Alberta Bonanni[1], Ilaria Zardo[3], Rinaldo Trotta[1*], and Armando Rastelli[1]*

[1] Institute of Semiconductor and Solid State Physics, Johannes Kepler University Linz, Altenbergerstrasse 69, A-4040 Linz, Austria

[2] Laser Processing Group, Instituto de Óptica, CSIC, C/Serrano 121, 28006 Madrid, Spain

[3] Department of Physics, University of Basel, Klingelbergstrasse 82, 4056, Basel, Switzerland

[4] Institute for Biophysics, Johannes Kepler University Linz, Gruberstrasse 40, A-4020 Linz, Austria

* e-mail: javier.martin_sanchez@jku.at ; rinaldo.trotta@jku.at

Two-dimensional transition-metal-dichalcogenide semiconductors have emerged as promising candidates for optoelectronic devices with unprecedented properties and ultra-compact performances. However atomically thin materials are highly sensitive to surrounding dielectric media, which imposes severe limitations to their practical applicability. Hence for their suitable integration into devices, the development of reliable encapsulation procedures that preserve their physical properties are required. Here, the excitonic photoluminescence of monolayers is assessed, at room temperature and 10 K, on mechanically exfoliated WSe₂ monolayer flakes encapsulated with SiO_x and Al_xO_y layers employing chemical and physical deposition techniques. Conformal flakes coating on untreated – non-functionalized – flakes is successfully demonstrated by all the techniques except for atomic layer deposition, where a cluster-like oxide coating is observed. No significant compositional or strain state changes in the flakes are detected upon encapsulation by any of the techniques. Remarkably, our results evidence that the flakes' optical emission is strongly influenced by the quality of the encapsulating oxide – stoichiometry –. When the encapsulation is carried out with slightly sub-stoichiometric oxides two remarkable phenomena are observed. First, there is a clear electrical doping of the monolayers that is revealed through a dominant trion – charged exciton – room-temperature photoluminescence. Second, a strong decrease of the monolayers optical emission is measured attributed to non-radiative recombination processes and/or carriers transfer from the flake to the oxide. Power- and temperature-dependent photoluminescence measurements further confirm that stoichiometric oxides obtained by physical deposition lead to a successful encapsulation, opening a promising route for the development of integrated two-dimensional devices.

Two-dimensional (2D) semiconductor transition-metal-dichalcogenides (TMDCs) present unique physical properties, such as a native sizeable bandgap, relatively large in-plane carrier mobility up to 250 cm$^2$/Vs, valley-dependent optoelectronics, strong light-matter interaction, and indirect-to-direct bandgap transition in monolayer crystals, which make them ideal candidates for ultra-compact optoelectronic, photonic and electronic applications [1-5]. Owing to strong spatial confinement of carriers and reduced dielectric screening of Coulomb interactions, electron and holes in monolayer TMDCs are tightly bound, thus forming exciton and multi-exciton complexes with binding energies up to 0.5-1 eV, more than one order of magnitude larger than in conventional III-V 2D quantum well nanostructures [6]. The simplest excitonic specie is the neutral exciton ($X^0$), which consists of an electron-hole pair bounded by Coulomb interaction. Singly charged excitons, or trions ($X^-$), quasi-particles consisting of two electrons (holes) and a hole (electron), have been reported [7,8] with binding energies up to 20-40 meV relative to $X^0$. Additionally, in WSe$_2$ monolayers, broad and relatively stable optical emission ~50 to ~200 meV below the $X^0$ transition is typically observed at low temperatures (< 60 K) and is associated with bound exciton states and other multi-exciton complexes [9-11].

The optical response of 2D semiconductor materials is therefore dominated by the radiative recombination of such exciton complexes that ultimately depends on many-body interactions of excitons and free carriers in the crystal. In addition, the large surface-to-volume ratio peculiar to these materials makes them highly sensitive to the surrounding environment. For example, organic solvents commonly used in well-established cleaning protocols, such as acetone or 2-propanol, can introduce unintentional doping in atomically-thin flakes [12], and the photoluminescence (PL) properties of doped TMDCs are strongly affected by physisorbed $O_2$, $O_3$ and/or $H_2O$ molecules, which tend to deplete the material of electrons, rendering the neutral exciton transition dominant over negatively charged exciton complexes [13]. The exciton dynamics in $WS_2$, $MoS_2$ and $WSe_2$ monolayer flakes is also affected by the substrate material, mainly due to strain, charge transfer or dielectric screening effects [3,14-16]. In the case of mechanically exfoliated flakes, these effects depend on the contact area between the exfoliated flakes and the substrate, which is affected by parameters such as substrate preparation procedure or stiffness of the tape/stamp used to transfer the flakes [17,18]. It is therefore not surprising to find discrepancies in published data on the characteristic optical emission features in $WSe_2$ monolayers at low temperatures due to uncontrollable experimental conditions. For instance, while some works report on well-defined neutral exciton, trion and localized states' emissions [9,10], broad bands with no clear peaks are found in others [11,19].

Besides fluctuations during preparation, strategies are needed to protect 2D materials against environmental modifications that could produce deleterious or uncontrollable effects on their electrical and optical properties. To this aim, different encapsulation techniques have been explored, such as embedding between inert h-BN [20] or dielectric materials, like $Al_2O_3$, $HfO_2$, $SiN_3$ or $SiO_2$ [21-25]. It is known that a conformal uniform oxide coating of 2D materials realized by chemical deposition techniques, such as atomic layer deposition (ALD), is challenging and requires a surface functionalization or pre-treatment step due to the absence of surface dangling bonds [24,26-28]. Most of the reports on $Al_2O_3$ [29], $HfO_2$ [23,30] or $SiN_3$ [31,32] encapsulation effects on TMDCs 2D materials, *e.g.* $MoS_2$ or $WSe_2$, are mainly focused on the study of their electrical response and rely on dielectric coating obtained by chemical deposition. Systematic studies on the effects produced by encapsulation on the flakes' optical response are scarce and there is controversy in their interpretation [33-35]. While in some works the optical emission modification of encapsulated $MoS_2$ flakes is attributed to the presence of residual strain or possible electrical charging [33,34], others relate it to dielectric screening effects [36]. In addition, possible material modifications that could affect the flake optical emission and the role of the employed deposition techniques are often neglected.

In this work, the optical properties of WSe$_2$ monolayers, encapsulated by chemical and physical deposition techniques with dielectric SiO$_x$ and Al$_x$O$_y$ layers of different thicknesses ranging from 5 to 30 nm, are systematically studied by PL spectroscopy. Special care is taken to prevent possible flake degradation or unintentional doping by avoiding any treatment procedure with organic solvents or further manipulation after the exfoliation. The composition, strain state and morphology of the encapsulated flakes are studied by x-ray photoelectron spectroscopy (XPS), Raman spectroscopy and atomic force microscopy (AFM), respectively. The stoichiometry of the encapsulating oxides is measured by XPS and correlated with the flakes' optical emission. Our results demonstrate that the stoichiometry of the oxide strongly influence the optical response of the encapsulated WSe$_2$. Dominant trion optical emission at room-temperature (RT) is found for flakes that were coated with sub-stoichiometric oxides, which indicates electrical doping of the flake due to the defective oxide. In contrast, no significant changes with respect to uncapped reference samples are found in the PL spectra when stoichiometric oxides are employed, indicating negligible charge transfer. Power-dependent PL at 10K and temperature-dependent PL measurements are performed on uncapped and encapsulated monolayers to further investigate the radiative recombination mechanisms. The suitability of each deposition technique to preserve the optical quality of encapsulated WSe$_2$ monolayer flakes is discussed.

**RESULTS AND DISCUSSION**

**Samples preparation, structural and compositional characterization of encapsulated WSe$_2$ monolayer flakes.** The samples used in this work are obtained by mechanical exfoliation of a bulk WSe$_2$ crystal (2dsemiconductors Inc.) on thermally oxidized Si substrates. The 280-nm thick SiO$_2$/Si layer presents a low surface roughness with a peak-to-peak value of about 1 nm. A set of samples consisting of uncapped (reference) and oxide-coated WSe$_2$ monolayer flakes with oxide thicknesses t=5, 10, 20 and 30 nm, and varying stoichiometry, is prepared. For the oxide deposition, we have chosen well-established and representative chemical deposition techniques – ALD and plasma-enhanced chemical vapor deposition (PECVD) – and physical deposition methods – electron-beam deposition (E-BEAM) and pulsed laser deposition (PLD) –.

The stoichiometry of the layers is determined by XPS and further details can be found in the supporting information, Figure S1 and Table S1. In particular, the physical deposition techniques led to the formation of stoichiometric $Al_2O_3$ oxides (E-BEAM and PLD). On the other hand, sub-stoichiometric oxide layers with oxygen depletion are obtained by chemical deposition: $Al_2O_{2.85}$ (ALD) and $SiO_{1.80}$ (PECVD). We would like to note that the deposition of the oxides used in this work has been performed employing standard experimental conditions under which oxide layers are routinely prepared in our laboratory. The idea behind our experiments is to unveil the effects of non-ideal conditions in terms of oxides stoichiometry or high energetic species, usually obtained in the PLD case, on the optical response of the encapsulated $WSe_2$ monolayers.

The surface morphology of as-exfoliated and oxide-coated $WSe_2$ is investigated by AFM. Figure 1a shows the AFM images of a $WSe_2$ monolayer flake coated by ALD for 50, 100 and 300 precursors [Trimethylaluminium (TMA)/$H_2O$] cycles, that correspond to a nominal oxide thickness $t_{nom}$=5, 10 and 30 nm of aluminium-oxide on the surrounding $SiO_2$ surface. Although pinhole-free coating is generally expected for ALD, conformal oxide growth requires the presence of uniformly distributed nucleation centers promoting the dissociation and reaction of the precursors. The AFM images and corresponding line-scans show that the growth is not conformal. Instead, nanometer-sized clusters are observed, indicating a three-dimensional deposition with peak-to-peak roughness values up to ~15 nm. Cluster coalescence is found to occur as the nominal thickness is increased up to 30 nm, eventually covering the whole flake surface. This ALD growth mode is observed on freshly exfoliated flakes, due to the low density of dangling bonds or impurities that could act as oxide nucleation centers on the monolayer's surface. This observation confirms the cleanness of the flake surface after the exfoliation and deposition procedures [26]. Interestingly, a conformal growth is observed when the flake encapsulation is carried out by PECVD, as shown in Figure 1b. The PECVD technique also involves chemical reactions during the oxide growth but, in this case, the dissociation of the precursors is enhanced by the plasma. The dissociated species react at the surface substrate and monolayers' surface to form a uniform oxide layer with no need of any nucleation center. On the other hand, in the physical deposition techniques employed here, a target material (oxide) is evaporated either by heating with an electron beam (E-BEAM) or by a pulsed laser (PLD). One of the main advantages of these techniques is that oxide layers with well controlled stoichiometry transfer from the target material are expected [37]. The oxide develops on the substrate surface and on the flake surface, by condensation. In principle, the adatom diffusivity on the substrate is determined by the kinetic energy of arriving evaporated species and no chemical reaction/precursor dissociation at the flake surface is required for the oxide to grow. Extra adatom mobility energy can be added by heating the substrate. In this work, E-BEAM deposition has been obtained for a substrate temperature of 200 ºC, whereas PLD has been performed with the substrate at RT. The results show uniform conformal encapsulation by both E-BEAM and PLD, as shown in Figure 1c-d. A similar surface roughness with a peak-to-peak roughness value of ~3 nm is found for all the samples, independently of the coating oxide thickness, except for the ALD case. A clear step of ~0.7 nm, corresponding to the monolayer thickness is measured in the profiles shown in Figure 1b-d.

To assess whether the flakes quality, composition or strain state are altered due to the encapsulation process, we have performed micro-Raman spectroscopy on individual monolayer flakes on selected samples. In particular, we focus on the vibrational modes $E_{2g}^1$ (in-plane) and $A_{1g}$ (out-of-plane), which are degenerate for monolayer $WSe_2$, giving rise to a single Raman peak at about ~ 250 cm$^{-1}$ [38]. This peak is particularly sensitive to strain [39] or doping in the flake [40]. Raman spectra collected in x(zz)x scattering geometry on uncapped and encapsulated flakes, with a capping oxide layer thickness of 30 nm, are shown in Figure 2a at wavenumbers around the $E_{2g}^1$/$A_{1g}$ modes. Solid lines represent fitting curves (the procedure followed in the analysis is described in the supporting information). No shift is observed after encapsulation within the system resolution (~0.5 cm$^{-1}$), indicating that, independent of the technique, the strain state and composition of the flakes are not appreciably altered by the coating. We obtain consistently similar results for different oxide thicknesses ranging from 5 to 30 nm. To further confirm our conclusions, XPS measurements on uncapped and 5-nm-thick oxide encapsulated flakes are compared. Figure 2b shows the XPS spectra for the 4f core-level signal of tungsten, where two main peaks attributed to the spin-orbit components W $4f_{5/2}$ and W $4f_{7/2}$ ($\Delta$ = 2.17 eV) are resolved. Most importantly, for all the samples but the one encapsulated by PLD, no $WO_x$ contribution to the overall spectra at higher binding energies is detected, which confirms that no flake oxidation takes place after encapsulation [27]. For PLD, it is found a small $WO_x$ component in the XPS spectrum that we attribute to mixing and swallow implantation of $Al_2O_3$ species at the flake during deposition inducing thus a slight oxidation [41,42]. By carefully analyzing the spectra for all the studied flakes, we find asymmetries of the W $4f_{5/2}$ and W $4f_{7/2}$ peaks, suggesting the presence of a second W $4f$ doublet shifted ~ 0.6 eV with respect to the main one (dashed lines in Figure 2b). These features have been previously reported and attributed to spatial variations in the Fermi level across the sample surface due to the presence of a mixture of van der Waals planes and non-van der Waals planes characterized by edge planes and stepped surfaces [43]. This is reasonable in our case, since the scanned area has a diameter around 400 μm, thus covering exfoliated flakes with different thicknesses and stepped surfaces. The different relative spectral weight of both doublets in different samples is associated with the uniqueness of each sample, which results from the fabrication process by mechanical exfoliation. No relative peak shifts are found in encapsulated flakes with respect to the uncapped flakes, ruling out significant compositional modification of the flakes upon encapsulation.

**PL on encapsulated WSe$_2$ monolayers at RT.** In order to study the optical emission of individual encapsulated flakes, we have first performed micro-PL measurements at RT. We emphasize that all the monolayers studied in this work feature the same emission at $E_{X^0}$=1.661 ± 0.005 eV before encapsulation, so that a direct comparison between them is possible. Figure 3a displays the normalized PL spectra for uncapped and encapsulated flakes. The black curve corresponds to the characteristic PL emission for an uncapped WSe$_2$ monolayer flake on a SiO$_2$/Si substrate and will be used as a reference. The spectrum is dominated by the neutral exciton (X$^0$) transition at $E_{X^0}$=1.661 eV, marked with a dashed line.

Remarkably, no significant PL peak shift or line-width broadening is found for WSe$_2$ monolayers encapsulated with stoichiometric Al$_2$O$_3$ oxides (E-BEAM and PLD). In agreement with this observation, previous studies demonstrate that dielectric screening effects on the optical transition energies of excitons in 2D materials surrounded by different dielectric media are negligible, since the expected exciton binding energy reduction is nearly compensated by a reduction of the free particle bandgap [44]. It is also remarkable that the slight oxidation induced by the PLD coating has not a dramatic effect on the RT optical response.

In contrast, for the flakes coated with sub-stoichiometric oxides $Al_2O_{2.85}$ (ALD) and $SiO_{1.80}$ (PECVD), we observe a significant PL emission red-shift up to around 35 meV with respect to the reference sample. Interestingly, for flakes coated by ALD, a gradual PL energy shift is observed as the oxide thickness is increased (Figure 3b). A similar total shift is found for PECVD encapsulation, but the shift is independent of the silicon-oxide layer thickness. The shifts are accompanied by a PL peak broadening: the full-width-at-half-maximum (FWHM) increases from ~40 meV for the uncapped sample up to ~60 and ~80 meV for PECVD and ALD encapsulated samples, respectively.

It is known that sub-stoichiometric oxides with oxygen vacancies, such as $SiO_{1.80}$ and $Al_2O_{2.85}$ used in this work, can act as sources of carriers for the underlying flake [45,46]. Moreover, additional localized states inside the band-gap are expected for defective amorphous oxides. Flake doping with electrons results in the gradual increase of the negative trion ($X^-$) emission compared to the neutral exciton, as reported for electrically gated 2D materials, such as $MoS_2$ [7], $MoSe_2$ [47], $WSe_2$ [8], or $WS_2$ [48].

When the carrier concentration in the flake increases due to the electrical doping produced by the sub-stoichiometric oxide, we expect a PL emission broadening due to the superposition of both neutral and charged components. This is exactly the case for the ALD encapsulated samples, where a gradual PL peak red-shift and broadening is observed by increasing the flake coverage, $i.e.$ electrical doping, as a consequence of the spectral weight change from neutral exciton to trion (Figure S3 in the supporting information). The maximum PL energy shift is ~35 meV, as expected for the trion relative binding energy in $WSe_2$ monolayers $E_{b,X^-}$~30 meV [8,49]. A similar total shift is found for PECVD encapsulated flakes. However, in this case, there is no smooth evolution, but rather a sharp change from a dominant neutral exciton to a dominant trion transition independently of the capping layer thickness. This is attributed to the uniform conformal flake encapsulation already achieved for the smallest investigated coverage of 5 nm.

The relative integrated PL intensity at RT before and after encapsulation ($I_{rel} = I_{PL_{capped}}/I_{PL_{uncapped}}$) for all encapsulated monolayers is depicted in Figure 4 as a function of the capping thickness. Average values $I_{rel} \sim 0.74 \pm 0.1$ and $I_{rel} \sim 0.59 \pm 0.1$ are found for monolayers encapsulated by E-BEAM and PLD, respectively. In the case of monolayers encapsulated with sub-stoichiometric oxides by ALD and PECVD, we would at first expect a similar PL intensity for the X⁻ peak compared to the X⁰ peak in undoped monolayers, following the results obtained by gate-controlled-doping [8,47]. In contrast, we find a significant PL decrease in doped monolayers. Since both Raman and XPS characterizations demonstrate no flake damage upon encapsulation (within the resolution of the techniques), the drop in PL intensity may be related to different mechanisms. One possibility is the development or enhancement of non-radiative recombination channels for the excitons or photogenerated carriers such as Auger processes, which are known to be particularly relevant for heavily-doped semiconductors. On the other hand, depletion of photogenerated holes or electrons through transfer mechanisms from the flake to traps present in the sub-stoichiometric oxide cannot be ruled out as a possible explanation for both PL decrease and electrical doping of the flake [50]. A qualitative correlation between oxide-induced doping and PL intensity decrease is observed for ALD deposition. In this case the PL intensity gradually decreases as the areal coverage of a flake (and hence the doping level) increases.

To further support the above interpretation about the gradual charging of the flake coated by ALD, Figure 1e shows the topography and surface potential maps measured by Kelvin Force Microscopy (KFM) [51] for a non-uniformly covered WSe₂ flake by ALD with $t_{nom}$=5 nm and a reference uncapped flake. For this experiment, a 10-nm-thick $Al_2O_{2.85}$ layer is deposited by ALD on the $SiO_2/Si$ substrate previous to the flakes transfer by mechanical exfoliation. The surface potential difference between the surrounding aluminum-oxide and the uncapped reference flake is $|\Delta V|$~200 mV. Interestingly, the partially covered flake presents an evident non-uniform surface potential distribution that closely replicates the topography of the coated flake with oxide clusters. It should be noted that the same total difference $|\Delta V|$~200 mV is measured between the uncovered parts and the surrounding aluminum-oxide similarly to the case of the uncapped flake. Most importantly, the uncovered and covered parts of the flake show potential fluctuations of about 100 mV. This observation, together with the PL data at RT reported above where an increasing electrical doping of the flake is revealed, suggests a local electrical doping of the flake at the nanoscale delimited by the size of the sub-stoichiometric oxide clusters. We point out here that the deliberate control of the electrical doping of selected areas in 2D materials is fundamental to define *p-n* junctions for applications in photovoltaics [52], light-emitting-diodes (LED) [53] or field-effect-transistors (FET) [54]. The manipulation of the electrical doping in monolayers has been demonstrated by different means, such as electrical gating [52,53], photoexcitation [55], self-limiting oxidation of native $WSe_2$ [56], photoinduced charge transfer [50,54,57], or surface functionalization [58]. Although electrical *n*-type doping on $WSe_2$ thin flakes was recently shown by encapsulation with sub-stoichiometric $SiN_x$ layers over micrometer-sized large areas [32], the results presented here at the nanoscale may find very interesting applications in nanodevices.

**Low temperature PL studies on encapsulated $WSe_2$ monolayers.** Figure 5a shows characteristic power-dependent PL spectra (at 10 K) for an uncapped $WSe_2$ monolayer. Figure 5b-e reports the spectra for flakes coated with 10 nm of oxide by the indicated deposition methods. We begin the discussion with the uncapped reference flakes. The overall PL emission consists of several peaks. The $X^0$ (1.747 eV) and $X^-$ (1.714 eV) can be well identified. There is some controversy regarding the origin of the emission at 1.696 eV. While it has been experimentally attributed to the biexciton complex (XX) formed by two bound excitons [10], recent calculations attribute this emission to an excited biexciton complex (XX*) with relative binding energy with respect to $X^0$ of about $E_{b,XX*}=59$ meV, whereas a binding energy $E_{b,XX}=20$ meV is predicted for XX emission [49,59,60]. Because of this ambiguity, we denote the transition at 1.696 eV as $P_1$. On the other hand, relative binding energies around 15 meV are predicted for exciton-trion ($X^0$-$X^-$) complexes, very close to the predicted value for XX [59]. We denote the weak PL emission centered at 1.732 eV ($E_b\sim15$ meV) with $P_0$. The PL peaks $L_0$, $L_1$, $L_2$ and $L_3$ are attributed to localized exciton states trapped in disordered potential wells that may be due to the underlying $SiO_2$ dielectric [61,62]. The multi-peak PL spectrum is fitted by Gaussian curves (Figure S3 in supporting information). We would like to note that very similar emission energies are found for all the transitions on monolayers exfoliated on several $SiO_2$/Si substrates, as shown in Figure S5a in supporting information. The relative binding energies depicted in Figure S5b for each of the transitions are comparable to other values reported throughout the literature. Although additional investigations are needed for a deeper understating of the origin of the PL transitions associated with localized states in the flakes, our observations suggest a stable relative binding energy for these transitions with some fluctuations, such as linewidth broadening. The latter might be attributed to an increase of the disorder in the confining potential wells due to the presence of defects in the flakes or local fluctuations of the underlying $SiO_2$/Si substrate roughness.

The dependence of the PL intensity of semiconductors on the excitation power can be often fitted with a power law $I \propto P^{\alpha}$ [63]. The integrated intensity for each of the peaks is calculated and fitted against the excitation power in a log-log plot to obtain the exponent $\alpha$ for uncapped and encapsulated flakes (Figure S6 and Table S3 in supporting information). In the uncapped flake, the peaks $X^0$, $P_0$, $X^-$ and $P_1$ show a nearly linear dependence with the excitation power ($\alpha \sim 1$) associated with the radiative recombination of free exciton complexes that follows a first-order kinetic [64]. On the other hand, the peaks $L_0$, $L_1$, $L_2$ and $L_3$ present a sub-linear dependence ($\alpha < 1$), which is attributed to bound exciton recombination in localized states [9]. Interestingly, a similar behavior is found for encapsulated samples with stoichiometric oxide by E-BEAM and PLD, as reported in Figure 5b-c. It should be noted that a broad defect-related bands are also observed for the encapsulated PLD sample labeled as $B_0$ and $B_1$ in Figure 5c, consistent with the XPS data shown above, attributed to flake oxidation and maybe to defect formation during deposition [42].

Encapsulated samples with sub-stoichiometric oxides by ALD and PECVD are notably different from the uncapped flakes (Figure 5d-e). Remarkably, free $X^0$ emission is not observed and mostly bound exciton emission dominates the overall PL spectrum at high excitation power. The energy position for PL peaks corresponding to $X^-$, $P_1$, $L_0$ and $L_1$ can be well fitted at similar energies with respect to the uncapped sample. In particular, a power law $\alpha > 1$ is found for these transitions. We attribute such a super-linear dependence to the presence of strong non-radiative decay channels, such as Auger processes, where the carrier recombination is no longer governed by a first-order kinetic [64]. This assumption is in line with the observed PL integrated intensity decrease at RT for encapsulated samples by ALD and PECVD shown in Figure 4. The observed broadening in the transitions can be due to an increase in the non-uniformities of the disorder potential upon encapsulation induced by the top oxide [61]. At the lowest excitation power values, only broad PL bands are observed: PECVD – $C_0$ (1.638 eV) and $C_1$ (1.593 eV) – and ALD – $D_0$ (1.638 eV) and $D_1$ (1.601 eV) –. The integrated PL intensity of these bands follows a linear power dependence and start to saturate at excitation power values around 200 µW (Figure S6 in supporting information). Since the quality of the flakes is well preserved upon encapsulation, as demonstrated by the XPS and Raman spectroscopy studies, such broad PL bands are likely due to the introduction of intra-gap energy levels upon encapsulation with the defective oxides and additional confining potential wells for excitons induced by the encapsulation with the oxide. Additionally, non-radiative recombination channels accounting for the super-linear power dependence of the $X^-$, $P_1$, $L_0$ and $L_1$ transitions are present, as discussed above.

Figure 6a-c shows temperature-dependent PL spectra for uncapped and encapsulated samples with stoichiometric oxides by E-BEAM and PLD. They are very similar and both show a blueshift for $X^0$, $P_0$, and $X^-$ transitions as the samples are cooled down from 300 K to 10 K, consistent with the Varshni law (Figure S7 in supporting information), which describes the temperature dependence of the band-gap in semiconductors:

$$E_g(T) = E_g(0) - \frac{\alpha' T^2}{(T+\beta)} \qquad (1)$$

where $E_g(0)$ is the energy bandgap at 0 K, while $\alpha'$ and $\beta$ are material related parameters. The fitted parameters are shown in Table S4 of the supporting information and are comparable to values previously reported for TMDCs [48]. At temperatures around 60 K, the bound exciton states $P_1$, $L_0$, $L_1$, $L_2$ and $L_3$ dissociate and consequently, the number of free $X^0$ and $X^-$ increases. The $X^-$ PL intensity decreases for temperatures above 100 K, being the $X^0$ transition dominant up to RT. Interestingly, we found that the emission related to localized states (bands $B_0$ and $B_1$ in Figure 6c) observed in the PLD sample persists up to temperatures as high as ~220 K, pointing at a different origin – *i.e.* flake partial oxidation – of the emission compared to the $P_1$, $L_0$, $L_1$, $L_2$ and $L_3$ transitions. The peak $P_0$ can be well resolved for the cases of the uncapped and E-BEAM encapsulated samples in the whole temperature range. However, it cannot be fitted for the PLD encapsulated sample, due to strong PL background induced by the broad PL band $B_0$ that extends from about 1.54 eV to nearly 1.70 eV. The peak $P_1$ is either quenching or merging with the $X^-$ emission for temperatures above 130 K.

In order to isolate and investigate the low-energy bands for encapsulated samples with sub-stoichiometric oxides by PECVD ($C_0$ and $C_1$) and ALD ($D_0$ and $D_1$), we have performed temperature-dependent PL experiments for a low excitation power of 10 μW. Under these experimental conditions, the low-energy broad bands dominate over the $X^-$ contribution, as shown in Figure 6d-e. As expected, the $X^-$ can be well fitted with the Varshni equation. The fitted parameters can be found in the supporting information. The broad PL bands undergo a redshift as the temperature increases. While the $C_1/D_1$ bands are not resolved above 160 K, the $C_0/D_0$ bands are stable up to RT, contributing to the emission broadening reported in Figure 3 and Figure S3. We therefore attribute the PL emission broadening observed at RT to an overall contribution of the $X^-$ emission and localized states induced by the encapsulation procedure. The nature of these states is clearly different from the other localized states discussed for uncapped and encapsulated flakes by E-BEAM/PLD, and might be ascribed to trapped excitons in disordered potential minima produced by the underlying substrate and/or the encapsulating oxide [61]. Additionally, such a disorder can be further enhanced by the spatially random electrostatic landscape provided by the sub-stoichiometric encapsulating oxide layer that remains charged upon carriers exchange with the flake.

**CONCLUSIONS**

In summary, we have performed a systematic study of the effects produced by oxide encapsulation on the photoluminescence of exfoliated $WSe_2$ monolayer flakes. We have shown that full conformal encapsulation on freshly exfoliated flakes can be successfully realized by PECVD, E-BEAM and PLD techniques. In contrast, oxide encapsulation by ALD results in inhomogeneous formation of clusters, which coalesce as the thickness is increased. The composition and strain state of the monolayers is well preserved upon encapsulation by all the techniques. Only encapsulated flakes by PLD show slight $WO_x$ traces detectable by XPS. Importantly, our results demonstrate that there is a high impact on the encapsulated flakes optical response depending on the deposited oxide stoichiometry. First, the RT PL properties of $WSe_2$ monoloayer flakes are preserved upon encapsulation with a stoichiometric oxide obtained by both physical deposition techniques E-BEAM and PLD. In contrast, the emission of flakes capped with sub-stoichiometric oxides by ALD and PECVD is dominated by trion emission, due to electrical doping through carriers exchange between the oxide and the flake. Additionally, a strong PL intensity decrease is also found, which suggests the presence of non-radiative mechanisms in the flake and/or a possible transfer of carriers from the flake to the oxide. Second, the power-dependent PL studies at 10K and temperature-dependent PL measurements show that encapsulated flakes with stoichiometric $Al_2O_3$ exhibit dominant free and bound exciton recombination processes similarly to uncapped flakes.

The flakes encapsulated by PLD display an additional broad band PL emission that might be due to the radiative recombination through defects originated in the flake by the arrival of energetic species during deposition. However, the flakes encapsulated with defective sub-stoichiometric oxides exhibit an emission consistent with the development of a high density of localized states, which leads to a broad PL band at low energies that is stable up to RT as discussed above. The origin of these localized states might be due to local potential wells fluctuations induced by the spatially random distribution of charges in the sub-stoichiometric oxide. From these results, we conclude that the most important factor for achieving a successful WSe$_2$ monolayer flakes encapsulation in terms of preserving their optical response is the oxide quality, *i.e.* the stoichiometry. As a consequence, all the deposition techniques used in this work have the potential to achieve a satisfactory oxide coverage of the 2D flakes, as long as the experimental deposition conditions are modified in order to achieve the correct stoichiometry and control the kinetic energy of the deposited species, in the PLD case. We believe that our findings can be extended to other 2D semiconductors and oxides/dielectrics, and that they will be of crucial relevance for the design and optimization of advanced optoelectronic devices where the design of complex semiconductor/dielectric nanostructures is mandatory. Moreover, the encapsulation procedure with stoichiometric oxides presented here provides a robust approach to protect 2D materials while preserving their optical properties. This will be particularly relevant for the transfer of oxide-nanomembranes embedding 2D materials on different target substrates, which can be for example accomplished using lift-off combined with dry transfer techniques. For instance, we envisage that these oxide membranes could be integrated onto micro-machined piezoelectric actuators to study with unprecedented detail the effect of large anisotropic strain fields on 2D semiconductors [65, 66].

## ACKNOWLEDGEMENTS

The authors would like to thank Georgios Katsaros and Tim Wehling for valuable discussions. Stephan Bräuer, Albin Schwarz and Ursula Kainz are acknowledged for technical support. A.M. acknowledges the financial support through BES-2013-062593. GG acknowledges support from FWF Project P 28018-B27. IZ acknowledges financial support from the Swiss National Science Foundation research grant (Project Grant No. 200021_165784). This work was partially funded by the Austrian Science Fund

**METHODS**

*WSe$_2$ monolayer flakes exfoliation:* Commercially available SiO$_2$ (280 nm)/Si substrates have been first degreased with acetone, 2-propanol and deionized water followed by oxygen plasma treatment for 10 minutes using a commercial system (200W) in order to enhance the flakes adhesion to the substrates [17]. The WSe$_2$ flakes are then transferred by Scotch mechanical exfoliation on the substrates where gold markers have been previously defined by optical lithography for flakes localization. Monolayer flakes are identified and localized with respect to the reference markers with an optical microscope. We limit the introduction of deleterious doping and/or contamination on the as-exfoliated flakes by avoiding any cleaning procedure with solvents. AFM and micro-PL are used on all the samples to confirm the monolayer thickness of the flakes. The monolayers used in this work present optical neutral exciton emission at similar energies of $1.661 \pm 0.005$ eV.

*Photoluminescence spectroscopy:* We use a standard confocal micro-PL setup equipped with a X-Y stage allowing spatial positioning of the laser spot with a resolution well below 1μm. A continuous-wave laser is used for excitation (λ=532 nm). The laser beam is focused on the sample by using a 50x objective with a numerical aperture value of NA=0.42. The temperature-dependent PL experiments are performed employing a He flow cryostat. PL is analyzed with a spectrometer with 750 mm focal length coupled to a Si CCD using a 150 grooves/mm grating. In order to perform the power-dependent measurements, we use a half-lambda wave plate combined with a fixed linear polarizer in the excitation line, which allows controlling reproducibly the laser excitation power.

*WSe$_2$ monolayer flakes oxide encapsulation by chemical and physical deposition. PECVD:* The samples were left in vacuum conditions (base pressure ~$1 \times 10^{-2}$ mbar) at 300 °C with a N$_2$ flux of 5 sccm for 30 minutes before deposition to desorb possible adsorbates from the surface. The silicon-oxide layers deposition is performed at 300 °C and 500 mbar by using SiH$_4$ (425 sccm) and N$_2$O (710 sccm) precursors with a plasma power of 10 W (13.56 MHz RF generator). The oxide layer thickness was controlled by the deposition time. *ALD:* The samples are left in vacuum conditions (base pressure ~$7x10^{-2}$ mbar) at 200 °C with a $N_2$ flux of 5 sccm for 30 minutes before deposition to desorb possible adsorbates from the surface. The aluminum-oxide deposition takes place at 200 °C by cycling Trimethylaluminium TMA (0.8 s)/$H_2O$ (1 s) precursors with a waiting time of about 8 seconds between pulses and $N_2$ flux of 20 sccm. The thickness of the oxide layers are controlled by the number of precursors' pulses. *E-BEAM:* The samples were annealed at 200 °C in vacuum (base pressure of ~$5x10^{-7}$ mbar) for 30 minutes to remove possible absorbents from the surface. The deposition is performed at 200 °C and $1x10^{-6}$ mbar while rotating the sample at 15 rpm. The oxide layer is deposited at a rate 2 Å/s. *PLD:* a home-built PLD system consisting of a UV laser ArF excimer ($\lambda = 193$ nm; 20 ns pulse duration) and a vacuum chamber equipped with a multi-target system is used. The laser beam is focused at an angle of incidence of approximately 45° onto a ceramic $Al_2O_3$ target. PLD deposition involves the formation of a laser-generated plasma formed by high kinetic energy species. The judicious choice of the deposition conditions allows to obtain good quality oxides for optical applications [37]. For this experiment, the samples are placed at 43 mm from the target surface and at 5 cm with respect to the center of the target normal, therefore using the so called off-axis deposition configuration that hinders the arrival of the higher kinetic energy species originated in the laser induced plasma plume and minimizes implantation and defect formation phenomena [67,68]. The oxide deposition is performed in vacuum (base pressure of $1x10^{-5}$ mbar) with the samples at RT using a laser energy density of $2.0 \pm 0.2$ J/cm$^2$ on the target with a repetition rate of 20 Hz.

*Raman spectroscopy:* Raman measurements are performed in backscattering geometry on isolated $WSe_2$ monolayers at RT, with excitation wavelength of 632.8 nm provided by a HeNe laser and power density of 14 kW/cm$^2$ (selected after checking that no heating effects are induced). The laser beam is focused with a 100x objective (NA=0.80). The scattered light is collected by a T64000 triple spectrometer equipped with 1800 g/mm gratings and liquid-nitrogen cooled multichannel charge couple device detector. Spectra are collected by selecting scattered light polarized either parallel or perpendicular to the polarization direction of incident light, which in the so-called Porto notation correspond to x(zz)-x and x(zy)-x, respectively.

*XPS spectroscopy:* The measurements are carried out with a Thermofisher Theta Probe XPS system, operated with a monochromatic Al-$K_\alpha$ x-ray beam at 15 kV, emission current of 6.7 mA and a spot size of 400 µm in diameter. All spectra are acquired in constant analyzer energy mode and constant pass energy of 50 eV, with FWHM=1.00 eV on Ag $3d_{5/2}$, and energy step size of 0.05 eV. The residual pressure of the ultra-high vacuum analysis chamber is held in the range of $10^{-8}$ mbar.

*KFM:* Kelvin Force Microscopy images [51] have been acquired at with the electrical excitation signal at the 2nd cantilever Eigenmode (f~426 kHz) [69]. A commercial AFM (Agilent 5600, USA) and conducting cantilevers (PtSi-FM, Nanoandmore, Germany) are used. The measured surface potential differences ΔV corresponds to the dc potential which nullifies the oscillation at 426 kHz. Assuming a parallel capacitor model the surface potential changes can be related to the surface charge density by σ =ΔV ε/h, where ε~$8 \times 10^{-11}$ F/m and h=5nm are the dielectric constant and the thickness of the oxide in Figure 5e.

*AFM:* A commercial Nanoscope system is used in tapping mode. Si AFM tips with resonance frequency around 300 KHz are employed. Data post-processing analysis is performed using the WSxM software [70].

**AUTHOR CONTRIBUTIONS**

JMS, RT and AR conceived the experiments. JMS supervised the work. JMS and AM fabricated the samples and performed the PL measurements and analysis of the data with the help of RT and AR. AM, RS and AH performed the oxide encapsulation of the flakes. ATML and AB performed the XPS characterization and data analysis. MDL and IZ performed the Raman spectroscopy measurements and data analysis. GG performed the Kelvin AFM probe measurements. JMS wrote the manuscript with inputs from all co-authors.

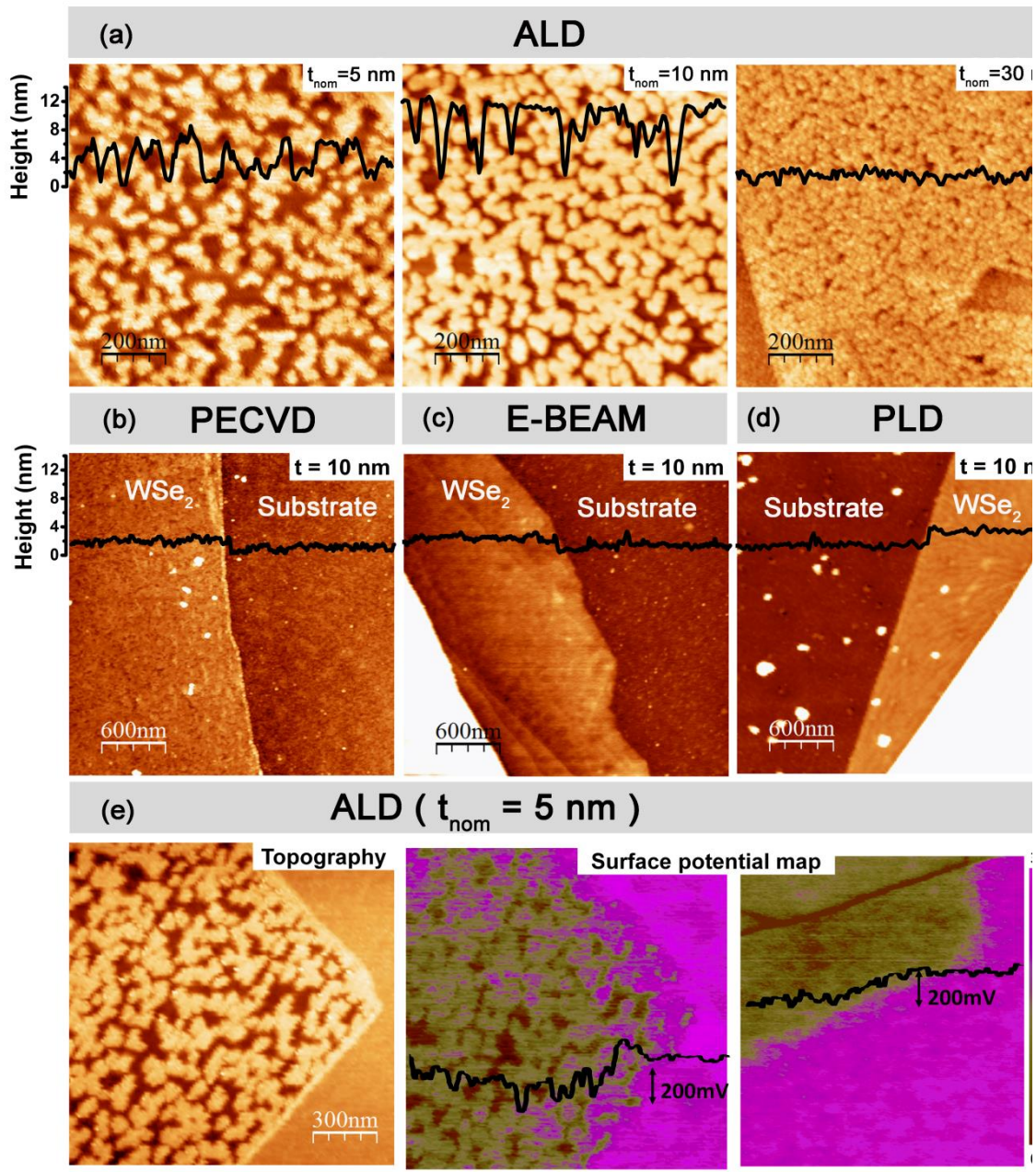

**Figure 1.-** (a) AFM pictures of encapsulated WSe$_2$ monolayer flakes by ALD (Al$_2$O$_{2.85}$) for a nominal oxide thickness t=5 nm, t=10 nm and t=30 nm. The line-scans over the AFM pictures are also plotted. The observed morphology corresponds to a cluster-like oxide coating of the flake due the lack of dangling bonds on the flakes' surface. (b-d) AFM pictures of encapsulated monolayers for an oxide thickness t=10 nm by PECVD (SiO$_{1.80}$), E-BEAM (Al$_2$O$_3$) and PLD (Al$_2$O$_3$). A uniform and conformal encapsulation is evidenced. Representative topography profiles are drawn over the images, where a step of about 0.7 nm corresponding to a monolayer thickness is measured for conformally encapsulated flakes.

Full oxide coating of the monolayer is observed for a thickness t=30 nm in the case of ALD. (e) Comparative topography and surface potential maps by KFM on a 5-nm-thick encapsulated monolayer by ALD and uncapped reference flake. A clear relation is found between the sub-stoichiometric $Al_2O_{2.85}$ oxide clusters and the static charges distribution over the monolayer.

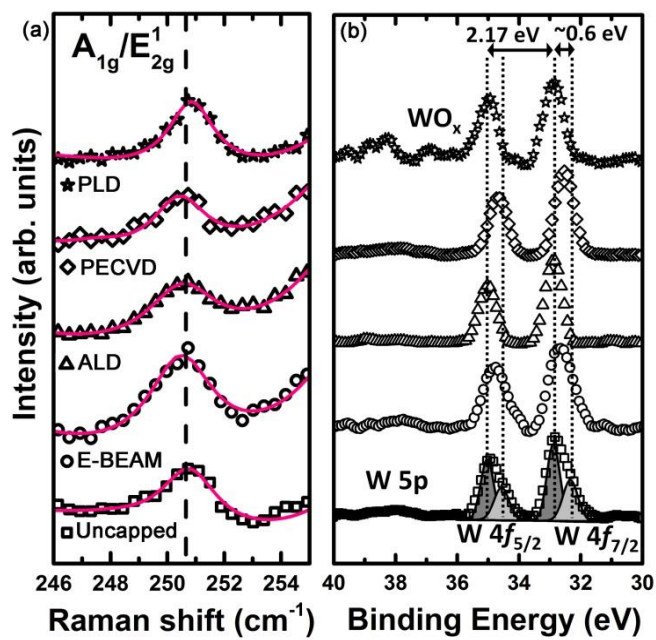

**Figure 2.-** (a) Stacked Raman spectra (open symbols) collected in the x(zz)-x scattering geometry on uncapped (squares) and encapsulated WSe₂ monolayers (t=30 nm) by E-BEAM (circles), ALD (triangles), PECVD (diamonds) and PLD (stars). The dashed line marks the frequency of the degenerate $E^1_{2g}$ and $A_{1g}$ modes distinctive of monolayers. Solid lines represent the best fits to the spectra. (b) XPS spectra of uncapped and encapsulated monolayers (t=5 nm) by E-BEAM, ALD, PECVD and PLD. The W 4f core-level signal with W $4f_{5/2}$ and $4f_{7/2}$ components are resolved together with a W 5p level contribution at a lower binding energy. An additional WOₓ component is found for the encapsulated flake by PLD, indicating partial flake oxidation.

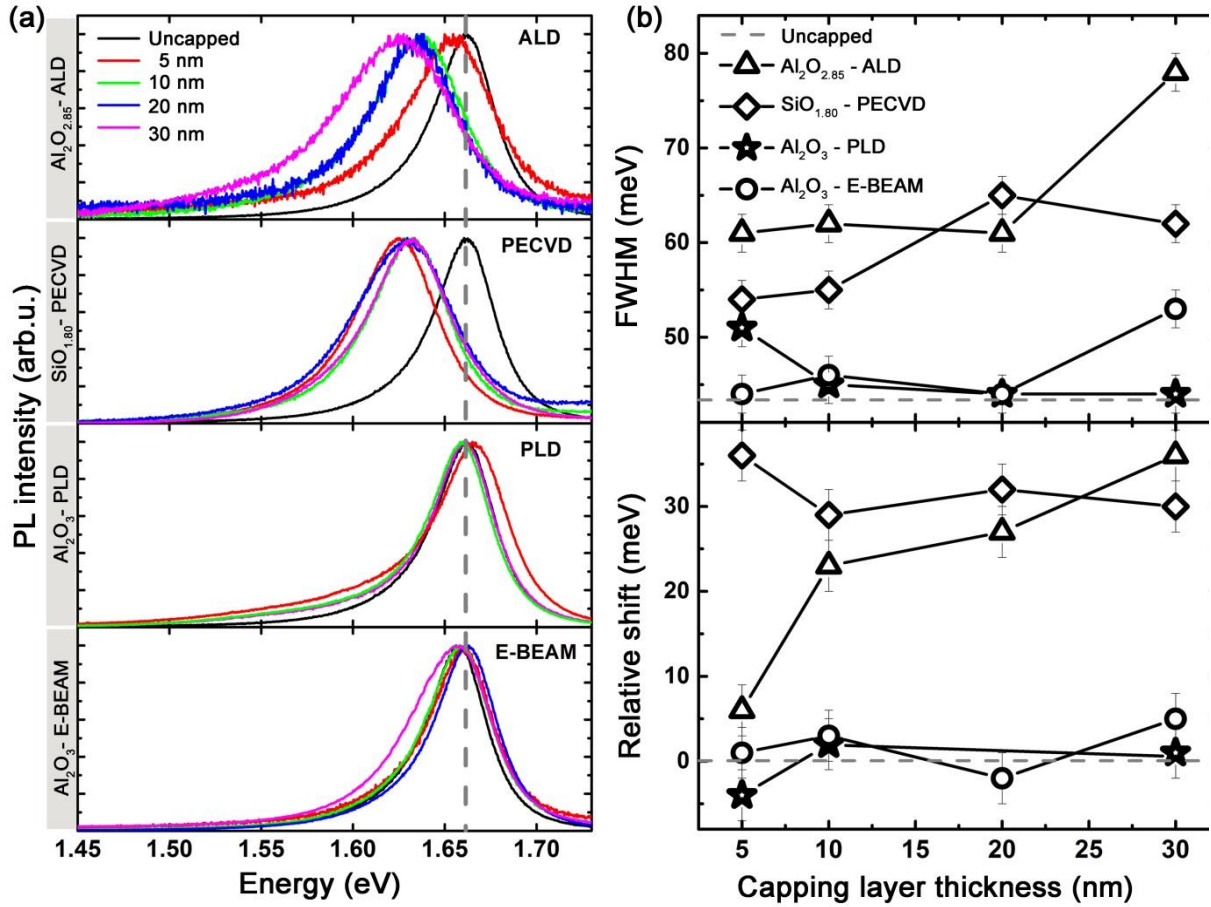

**Figure 3.-** (a) Normalized micro-PL spectra at RT of uncapped (black curve) and encapsulated $WSe_2$ monolayers for different capping layer thicknesses and varying oxide stoichiometry. Oxides are produced by ALD ($Al_2O_{2.85}$), PECVD ($SiO_{1.80}$), PLD ($Al_2O_3$) and E-BEAM ($Al_2O_3$). (b) Corresponding FWHM and relative PL peak energy shift values for all the samples. The values for the uncapped reference monolayer are marked with a horizontal dashed line.

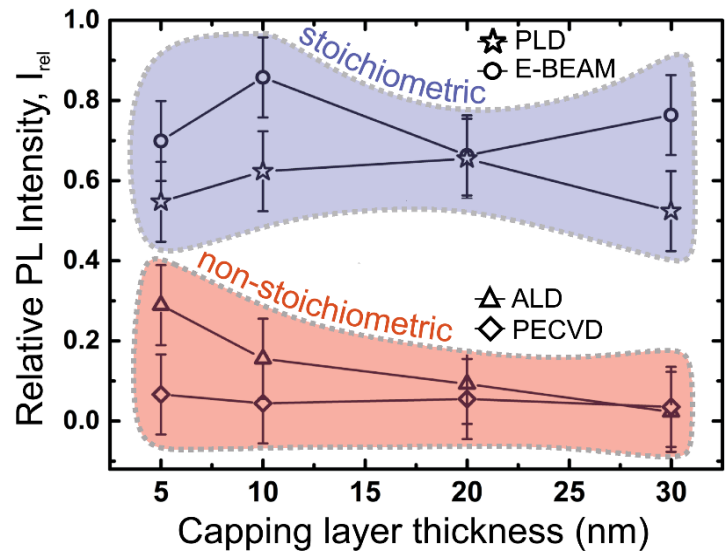

**Figure 4.-** Relative integrated PL intensity at RT of encapsulated WSe$_2$ monolayers for different capping layer thicknesses (t=5, 10, 20 and 30 nm) with respect to uncapped monolayers ($I_{rel} = I_{PL_{capped}}/I_{PL_{uncapped}}$). Oxides are produced by ALD (Al$_2$O$_{2.85}$), PECVD (SiO$_{1.80}$), PLD (Al$_2$O$_3$) and E-BEAM (Al$_2$O$_3$). An excitation power of 10 µW is used.

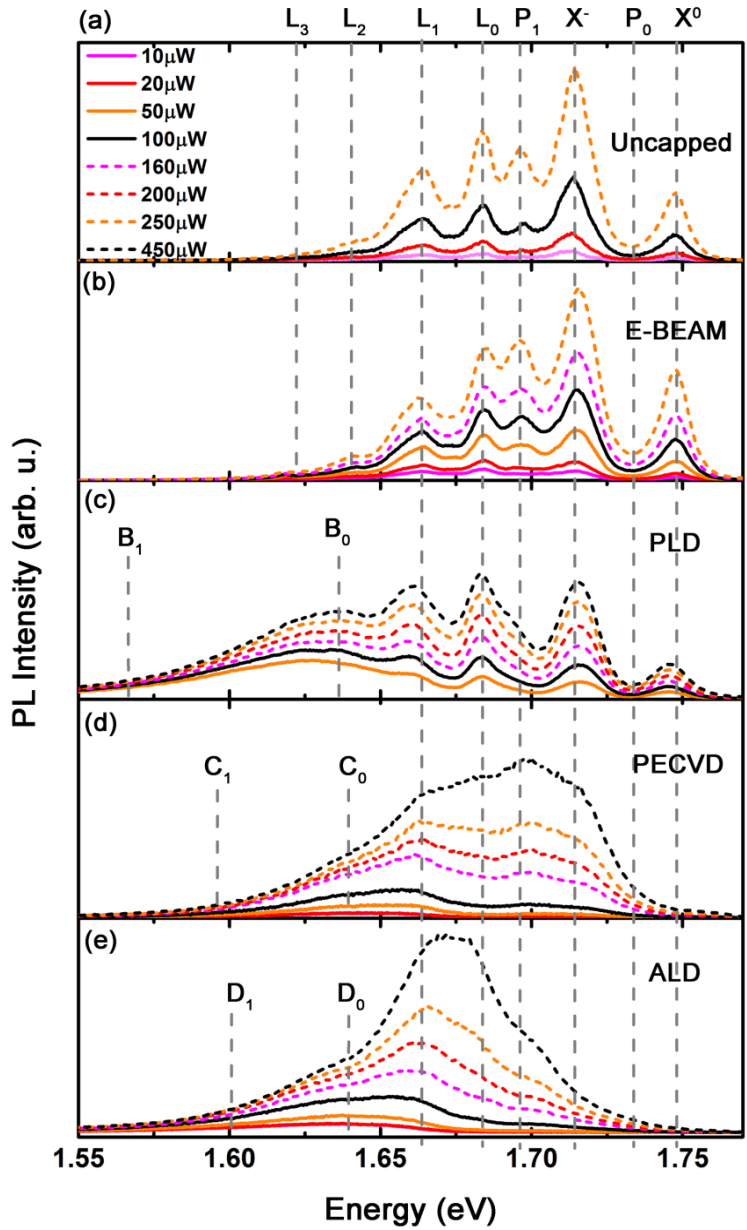

**Figure 5.-** Power-dependent PL spectra at 10 K for an uncapped WSe$_2$ monolayer (a), and encapsulated (t=10 nm) monolayers by E-BEAM (b), PLD (c), PECVD (d) and ALD (e). The energy position for X$^0$, P$_0$, P$_1$ and X$^-$ free exciton, and L$_0$, L$_1$, L$_2$ and L$_3$ bound exciton recombinations are marked with dashed vertical lines. Additional PL bands are indicated for encapsulated monolayers by PLD (B$_0$ and B$_1$), PECVD (C$_0$ and C$_1$) and ALD (D$_0$ and D$_1$).

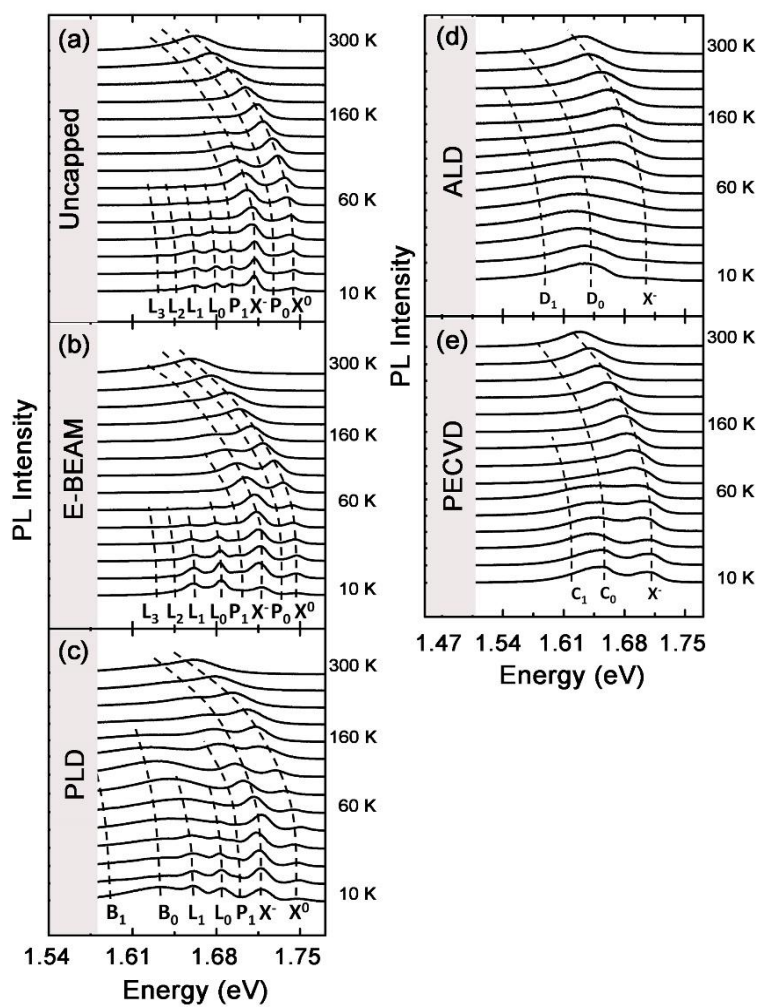

**Figure 6.-** Temperature-dependent PL spectra at an excitation power of 10 µW, for uncapped (a) and encapsulated WSe$_2$ monolayers with stoichiometric oxides by E-BEAM (b) and PLD (c). The temperature-dependent PL spectra for encapsulated monolayers with sub-stoichiometric oxides by ALD and PECVD are shown in (d) and (e), respectively. The dashed lines serve as a guide for the eye to follow the energy shift of the PL components with temperature.